# Inorganic Tin Perovskites with Tunable Conductivity Enabled by Organic Modifiers

Md Azimul Haque[1], Tong Zhu[2], Luis Huerta Hernandez[1], Roba Tounesi[1], Craig Combe[1], Bambar Davaasuren[3], Abdul-Hamid Emwas[3], F. Pelayo García de Arquer[4], Edward H. Sargent[2], Derya Baran[1*]

[1]*King Abdullah University of Science and Technology (KAUST), Physical Science and Engineering Division (PSE), KAUST Solar Center (KSC), Thuwal 23955-6900, Saudi Arabia*

[2]*Department of Electrical and Computer Engineering, University of Toronto, 35 St George Street, Toronto, ON M5S 1A4, Canada*

[3]*King Abdullah University of Science and Technology (KAUST), Corelabs, Thuwal 23955-6900, Saudi Arabia*

[4]*ICFO – Institut de Ciències Fotòniques, The Barcelona Institute of Science and Technology, Barcelona, Spain*

[*]*Email: derya.baran@kaust.edu.sa*

**Abstract**

Achieving control over the transport properties of charge-carriers is a crucial aspect of realizing high-performance electronic materials. In metal-halide perovskites, which offer convenient manufacturing traits and tunability for certain optoelectronic applications, this is challenging: The perovskite structure itself, poses fundamental limits to maximum dopant incorporation. Here, we demonstrate an organic modifier incorporation strategy capable of modulating the electronic density of states in halide tin perovskites without altering the perovskite lattice, in a similar fashion to substitutional doping in traditional semiconductors. By incorporating organic small molecules and conjugated polymers into cesium tin iodide ($CsSnI_3$) perovskites, we achieve carrier density tunability over 2.7 decades, transition from a semiconducting to a metallic nature, and high electrical conductivity exceeding 200 S/cm. We leverage these tunable and enhanced electronic properties to achieve a thin-film, lead free, thermoelectric material with a near room-temperature figure-of-merit (ZT) of 0.21, the highest amongst all halide perovskite thermoelectrics. Our strategy provides an additional degree of freedom in the design of halide perovskites for optoelectronic and energy applications.

**Introduction**

Over the last decade, halide perovskites have attracted interest for electronic applications, including photovoltaics, LEDs, transistors, and radiation detectors due to their diversity of function, tunability, and solution processing.[1-4] Early reports on halide perovskites were focused on lead (Pb)-based materials and devices motivated by their promising performances in photovoltaic applications. Lately, tin (Sn) and Sn-Pb mix perovskites emerged due to several drawbacks of Pb-perovskites such as toxicity and limited band gap range that would not be ideal for tandem photovoltaics and thermoelectrics.[5-7] Although there has been continuous increase in reports on material properties and characteristics related to Sn perovskite films and devices, the understanding and mastery of electronic properties that determine device performance, such as charge transport, mobility, carrier density, and doping is still in its early stages. Sn perovskites demonstrate tunability in optoelectronics, but a wider doping range is imperative for applications such as thermoelectrics. Despite previous attempts at substitutional and molecular doping, the conductivity values after doping did not reach high levels due to the initial low carrier density or conductivity of the Sn perovskites under investigation.[8-11] The difficulty in doping halide perovskite semiconductors stems from their intrinsic defects, which exhibit strong charge compensating nature.[12] Understanding the underlying mechanism of perovskite-dopant (additive) interaction can enhance their conductivity, potentially opening up new applications beyond photovoltaics. Conductivity is primarily governed by carrier density and mobility. Simultaneously improving both can serve as another potential route to enhance conductivity.

The all-inorganic perovskite $CsSnI_3$ possess rich electronic properties, expanding the functionality of this class of materials to transistors and thermoelectrics.[13, 14] The ionic-covalent nature of $CsSnI_3$ enables the coordination of functional groups from organic molecules, providing a unique opportunity for organic modification.[15] However, the organic modification of $CsSnI_3$ has been largely limited to reducing organic additives to enhance photovoltaic performance, with little exploration of the vast library of organic semiconductors. This is due to the lack of general guidelines on how organic molecules interact with $CsSnI_3$ lattice. Specifically, difficulty in tuning the electrical conductivity of $CsSnI_3$ arises from its high intrinsic carrier density and the multiple oxidation states of Sn. Conjugated organic molecules as additives have been shown to modify the charge transport properties and reconfigure band edge states in Pb perovskites.[7, 16, 17] This approach could be extended to Sn perovskites to achieve a tailored density of states (DOS) that meets the requirements of high conductivity for applications such as thermoelectrics. The major challenge in designing such a material system is identifying organic molecules that can contribute to the Sn perovskite DOS near the band edges.

This study presents a series of organic modifiers to modulate the electronic properties of $CsSnI_3$. Incorporation of organic components into the inorganic $CsSnI_3$ matrix leads to binding interaction through functional groups, influencing the charge transport. We demonstrate selective semiconducting and metallic transport in these $CsSnI_3$ hybrids (organic modifier added $CsSnI_3$) enabled by various organic small molecules and polymers. Through tunability of carrier density and mobility, along with processing optimizations, we achieve an enhancement of electrical conductivity from 78 S/cm to 216 S/cm at room temperature, which is among the highest for solution-processed halide perovskites. Density functional theory (DFT) calculations reveal strong interactions between the organic molecules and $CsSnI_3$, giving rise to new features in the DOS. The position of the DOS contribution by organic moieties in the band structure influences the overall electronic properties. The enhanced transport properties of modified $CsSnI_3$ exhibit superior thermoelectric performance with a near-room temperature ZT of 0.21.

**Results**

Commonly used additives or treatments for Sn perovskite photovoltaics predominantly reduce the electrical conductivity of $CsSnI_3$ (Figure 1a) which is counterproductive for applications requiring high conductivity. We started our investigation by introducing a benzoquinone-based small molecule, 2,5-dihydroxy-1,4-benzoquinone (DB) in the $CsSnI_3$ precursor. Benzoquinones are redox-active organic molecules with diverse applications, previously employed for redox-couple in dye-sensitized solar cells, elimination of sub bandgap states in quantum dots, and improving morphology in Pb perovskites.[18-20] DB-containing $CsSnI_3$ films exhibit metallic behavior (decreasing electrical conductivity with temperature) in contrast to semiconducting behavior of pristine $CsSnI_3$ (Figure 1b). DFT calculations elucidate that DB interacts with the $CsSnI_3$ lattice (Figure 1c) and contributes to the DOS near the conduction band maximum region (Figure 1d). In order to gain control over the position of DOS, we explored a series of organic small molecules and polymers as modifiers. Similar to DB, two other small molecules: Catechol (CT) and phenazine-2,3-diol (PD) which contains an electron-deficient aromatic component, were investigated. We also explored Polyethylene glycol (PEG), which has been found to impart enhanced moisture stability in Pb-perovskites.[21] Finally, we examined polymers such as Poly(3-

hexylthiophene-2,5-diyl) (P3HT), Polyaniline (PANI), Poly(3,4-ethylenedioxythiophene) polystyrene sulfonate (PEDOT:PSS), and Poly(3,4-ethylenedioxythiophene):bis-poly(ethyleneglycol) (PEDOT:PEG) block co-polymer to determine their effect on DOS and assess the feasibility of incorporating polymers into $CsSnI_3$ films.

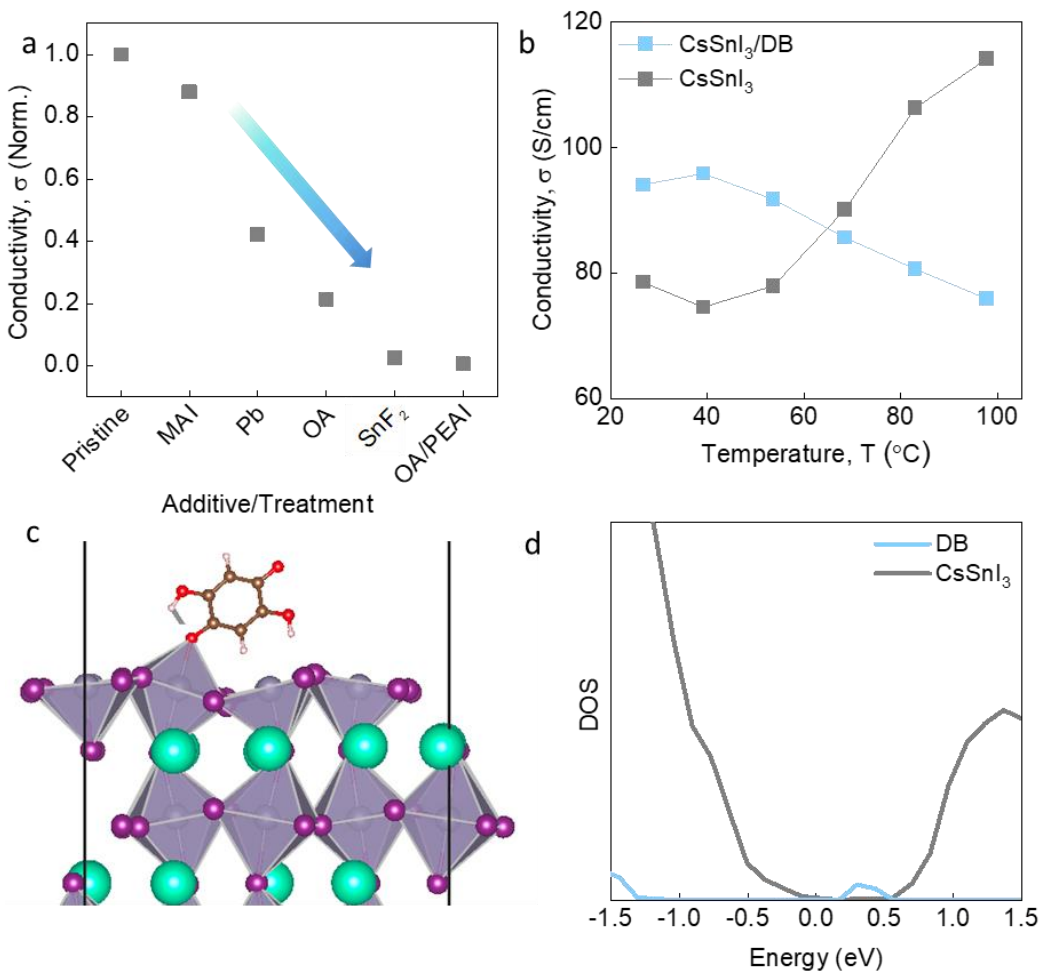

**Fig. 1 Effect of organic modifier (DB) on $CsSnI_3$. a** Electrical conductivity of pristine $CsSnI_3$ and with various additives. **b** Electrical conductivity of pristine $CsSnI_3$ and $CsSnI_3$/DB. **c** PBE+TS relaxed slab structure for DB molecule and $CsSnI_3$. **d** PBE+SOC calculated projected DOS for DB added $CsSnI_3$. The DB and $CsSnI_3$ contributions are shown as blue and gray color, respectively.

The addition of organic modifiers preserves the perovskite structure of $CsSnI_3$ but has a marked effect on the morphology of the films (Figure S1-S2). PEG addition resulted in an inferior microstructure with extensive pinholes, while P3HT and DB show a morphology with smaller grains. The modifiers PEDOT:PSS, PANI, and PEDOT:PEG were visible on the film surface. In terms of electronic properties, $CsSnI_3$ films with P3HT, PD, PEG, CT demonstrate semiconducting (increase in electrical conductivity with temperature) behavior, whereas those with PEDOT:PSS,

PANI, DB and PEDOT:PEG show metallic trend (decrease in electrical conductivity with temperature) as shown in Figure 2a. Among the Catechol inspired small molecules, only DB, which has bidentate ligands, contributes to DOS near band edge while CT and PD contribution are far from valence band, indicating certain design rules for organic moieties affecting the DOS in CsSnI₃ hybrids. The organic modifiers offer on-demand tunability of semiconducting or metallic behavior and significantly improve the conductivity of CsSnI₃ (Figure S3a). The room-temperature conductivity of CsSnI₃ hybrids ranges widely from 1.7 to 156 S/cm as a function of the organic modifier. Temperature-dependent Seebeck coefficient ($S$) confirms that all CsSnI₃ hybrids maintained the p-type nature of CsSnI₃ (Figure S3b).

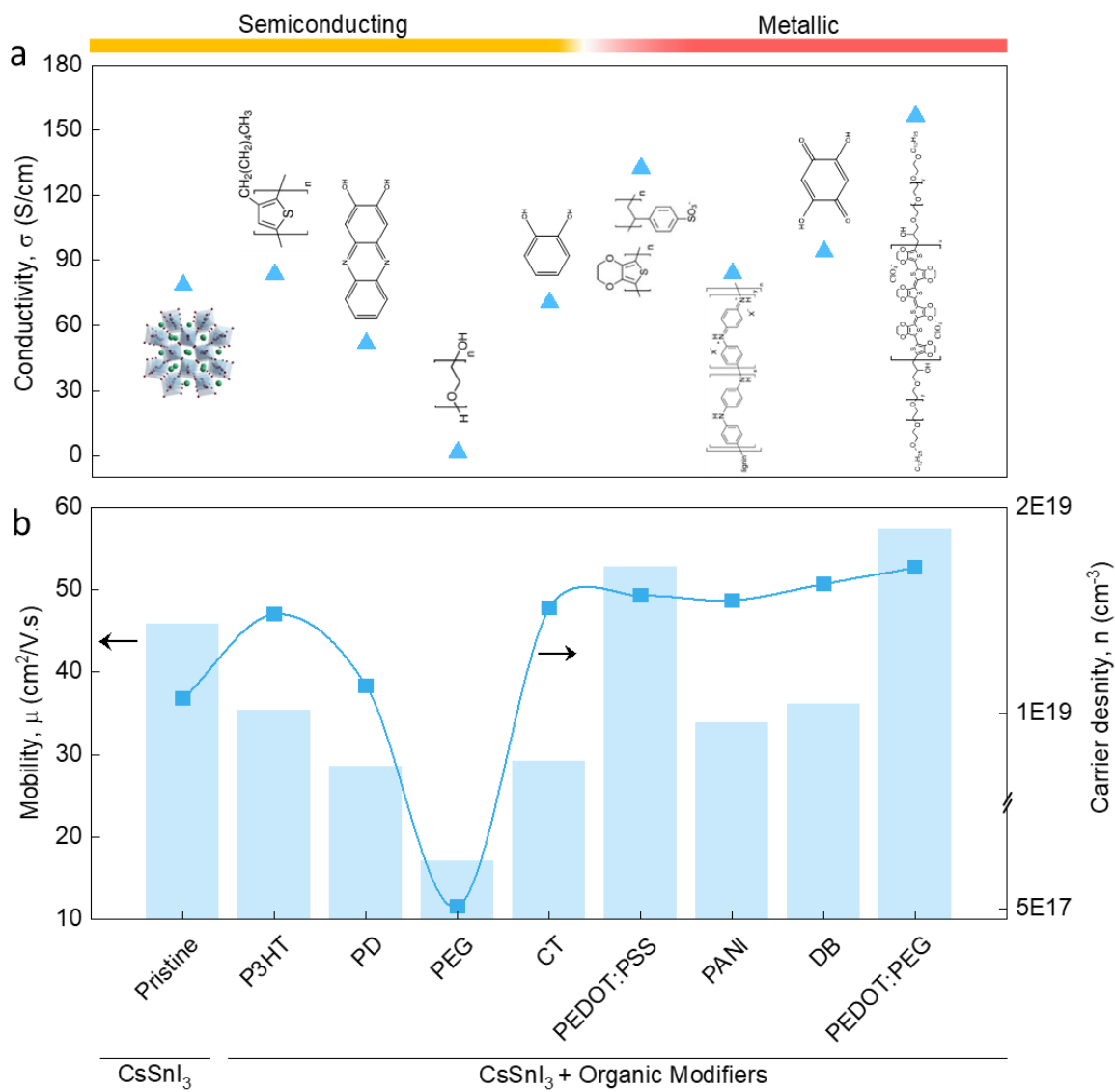

**Fig. 2 Effect of organic modifiers on CsSnI₃ electrical properties.** Room-temperature **a** electrical conductivity **b** Mobility and carrier density of pristine CsSnI₃ and with various organic

modifiers. Carrier density axis in b has been custom labelled. $CsSnI_3$/PEDOT:PSS and $CsSnI_3$/PEDOT:PEG show simultaneous enhancement in mobility and carrier density compared to pristine $CsSnI_3$. Loading amount of organic modifiers in $CsSnI_3$ precursor was 1 mg/ml for PD, PEG, CT, PANI, DB and 100 µl for PEDOT:PSS and PEDOT:PEG. For $CsSnI_3$/P3HT, 1 mg/ml P3HT in chlorobenzene was used as antisolvent during $CsSnI_3$ spin coating process. Loading amount optimization is shown in Figure S4.

To further understand the semiconducting and metallic behavior, we measured the carrier densities at room temperature. The modifiers with metallic behaviors (PANI, DB, PEDOT:PSS, PEDOT:PEG) show marginally increased carrier density compared to those of semiconducting behavior modifiers (P3HT, PD, PEG, CT) and pure $CsSnI_3$ (Figure 2b, S5). Interestingly, higher mobility was observed for PEDOT:PSS/$CsSnI_3$ and PEDOT:PEG/$CsSnI_3$ hybrids compared to pristine $CsSnI_3$. In addition, $CsSnI_3$/PEDOT:PEG shows 25 percent higher mobility than pristine $CsSnI_3$ (Figure 2b). $CsSnI_3$/PEG shows the lowest carrier density and mobility (due to insignificant DOS contribution by PEG and inferior film quality), whereas $CsSnI_3$/PEDOT:PEG exhibited the highest mobility and carrier concertation leading to high conductivity. Moreover, PEDOT:PEG improved the air stability of $CsSnI_3$ which can be attributed to PEG (Figure S6).

DFT calculations were conducted to investigate the mechanism underlying the different electrical transport behaviors (metallic or semiconducting) of $CsSnI_3$ with various organic modifiers. The (PBE+TS) relaxed structures within the lowest total energy for eight functional group molecules are shown in Figure 3a. In the band structure of all the polymers and molecules showing semiconducting behavior (P3HT, PD, PEG, CT), no significant changes in the band edge region (defined as VBM-0.5 eV ~ CBM) were observed compared to the pure $CsSnI_3$ slab (Figure S7). Conversely, modifiers PANI, DB, PEDOT (in PEDOT:PSS and PEDOT:PEG) exhibiting metallic behavior presented obvious changes in the band edge regions (Figure 3b), leading to a higher probability of more free carrier density, in contrast to functional groups with no band edge contribution (semiconducting behavior) which is consistent with experimental charge carrier density measurements. The observed different semiconducting and metallic behaviors can be explained by the strong interactions of the polymers and molecules with the $CsSnI_3$ surface in the grain boundary area, which may lead to different changes in the electronic structures and affect their free carrier densities. For functional group molecules that contribute to the band edge region in the DOS (PANI, DB, PEDOT), electrons coming from added polymers and molecules have a greater probability of transferring from the valence band to the conduction band, thus offering more free carriers (free electrons and holes) with increasing temperature. Higher free carrier densities at increased temperature led to metallic behavior in the temperature-dependent conductivity for PANI, DB, PEDOT:PSS, and PEDOT:PEG. For molecules with no contributions to the band edge region in the DOS (P3HT, PD, PEG, CT, PSS) show rare probability for electrons of the molecules to transfer from the valance band to the conduction band with increasing temperature, leading to the semiconducting behavior similar to pristine CsSnI$_3$. Supplementary note 1 (Figure S8-S9) provides further discussion on these findings.

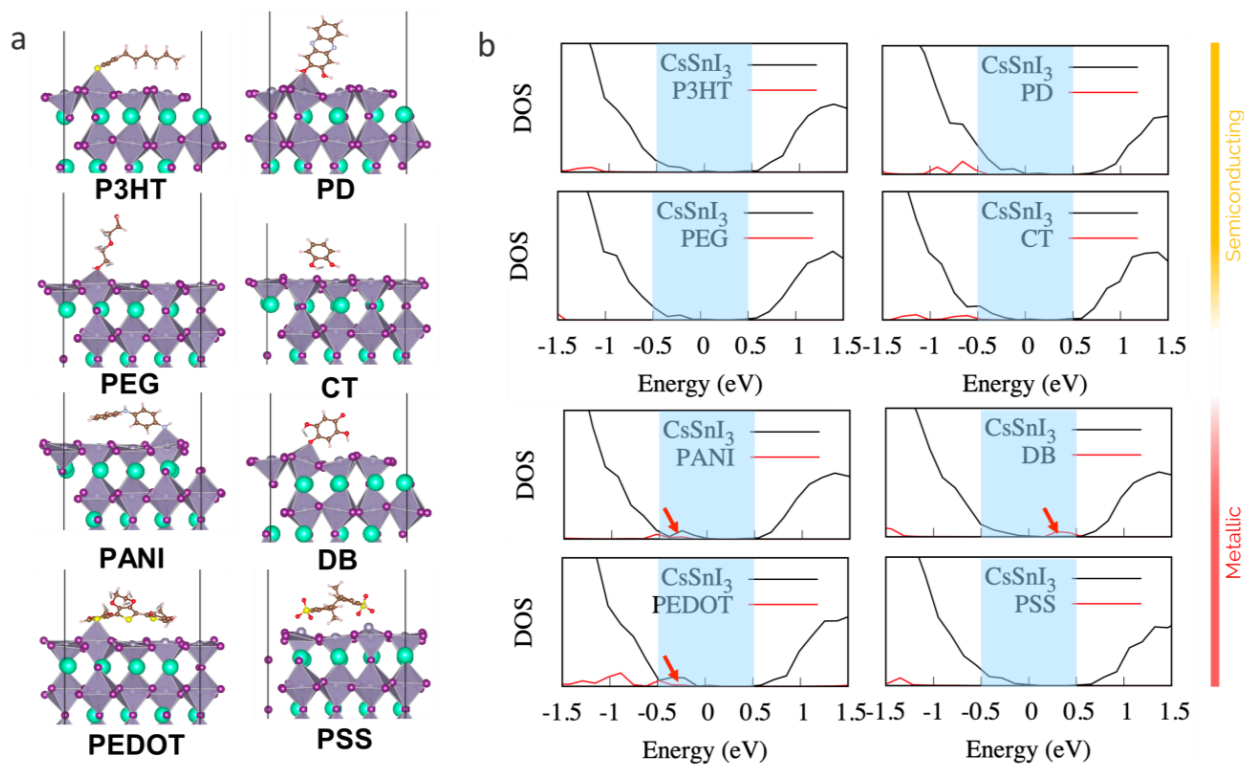

**Fig. 3 DFT calculations of organic modifier interaction and DOS changes. a** PBE+TS relaxed slab structure for the eight selected functional group molecules. **b** PBE+SOC calculated the projected density of states (DOS) for the eight selected functional groups molecules. While the contribution from the CsSnI$_3$ is summarized and labeled as a black line, the contribution from different functional groups is labeled as a red line. The band edge regions (VBM-0.5 eV ~ CBM) are labeled as shadowed blue areas. The functional group molecules, which have contributions in this band edge regions are pointed by the red arrow.

## Discussion

Inspired by the performance and stability enhancement of CsSnI$_3$ imparted by PEDOT:PEG, we worked on further improving the electronic properties through process engineering. The optimization details for CsSnI$_3$/PEDOT:PEG are discussed in supplementary note 2 (Figure S10-S14). Pristine $CsSnI_3$ and $CsSnI_3$/PEDOT:PEG precursor solutions were heated at 70 °C and casted while it was hot. Compared to room temperature (RT), hot solutions (H) demonstrated significantly enhanced electrical conductivity due to increased carrier density ($n$) and mobility ($\mu$) (Figure 4a, S15). Carrier density (mobility) increased from $1.07 \times 10^{19}$ cm$^{-3}$ (45.8 cm$^2$V$^{-1}$s$^{-1}$) to $1.82 \times 10^{19}$ cm$^{-3}$ (74.3 cm$^2$V$^{-1}$s$^{-1}$) for $CsSnI_3$(RT) and $CsSnI_3$/PEDOT:PEG(H) films, respectively. The present electrical conductivity is among the highest for solution-processed halide perovskite films (Figure S16). Consequently, the Seebeck coefficient decreased for hot-cast films due to increased conductivity (Figure S17). To confirm that this improvement was not due to Sn oxidation, we measured the NMR spectra of the precursor solutions and did not observe any signs of oxidation (Supplementary note 3, Figure S18). Furthermore, intentionally oxidizing the precursor solution resulted in films with lower mobility due to high carrier density (Figure S19). These observations reassure the conductivity enhancement was not due to Sn oxidation.

Compared to RT films, there was a crystallinity increase in the case of hot cast films, which explains the improved mobility (Figure 4b). As orientation and crystallinity can affect the electronic transport, we checked for orientation changes by 2D XRD and found no variations between RT and hot cast films (Figure 4c). Temperature-dependent in-situ XRD demonstrated no change in the phase transition (orthorhombic to tetragonal) temperature of $CsSnI_3$ after PEDOT:PEG addition or heating, ruling out any possible effect on charge transport (Figure 4d-e, S20-21). To summarize, enhanced carrier density and mobility along with improved crystallinity contribute to the high electrical conductivity of hot-cast $CsSnI_3$/PEDOT:PEG hybrid films. A power factor, $\sigma S^2$ (PF) of above 150 $\mu$W/mK$^2$ was achieved for PEDOT:PEG/$CsSnI_3$(H) owing to the high electrical conductivity of hot-cast films, which is significantly higher than the PF of $CsSnI_3$(RT) (Figure 4f). To evaluate the thermoelectric figure of merit ($ZT$) of $CsSnI_3$(RT) and PEDOT:PEG/$CsSnI_3$(H), we measured the in-plane thermal conductivity by a chip-based 3$\omega$ technique (Figure S22a). The thermal conductivity ($\kappa$) of $CsSnI_3$/PEDOT-PEG(H) is lower than that of pristine $CsSnI_3$, which can be attributed to enhanced phonon scattering due to organic additives.[22] This resulted in a ZT of 0.15 at RT and a best ZT of 0.21 at 53 °C (Figure S22b) for $CsSnI_3$/PEDOT-PEG(H).

$CsSnI_3$ undergoes structural changes at room temperature upon air exposure. 3D $CsSnI_3$ becomes metastable $Cs_2Sn_2I_6$, which is 1D in nature and further translates to air-stable 0D $Cs_2SnI_6$ (Figure S23). All the phases are crystalline, with $CsSnI_3$ and $Cs_2SnI_6$ black in color while $Cs_2Sn_2I_6$ is yellow (Figure S24). Interestingly, CsSnI₃ is p-type, and Cs₂SnI₆ is n-type, as evident from the sign of their Seebeck coefficient (Figure S25). We also evaluated the thermoelectric properties of low dimensional n-type Cs₂SnI₆, and a best ZT value of 0.038 was obtained at 97 °C owing to its ultralow thermal conductivity (Supplementary note 4, Figure S26-27). Table 1 summarizes the thermoelectric parameters of CsSnI₃(RT), CsSnI₃/PEDOT:PEG(H), and Cs₂SnI₆ films for temperatures corresponding to the highest ZT. As a result of high electrical conductivity and ultralow thermal conductivity, CsSnI₃/PEDOT:PEG represents the highest ZT so far for perovskite thermoelectrics (Figure S28).

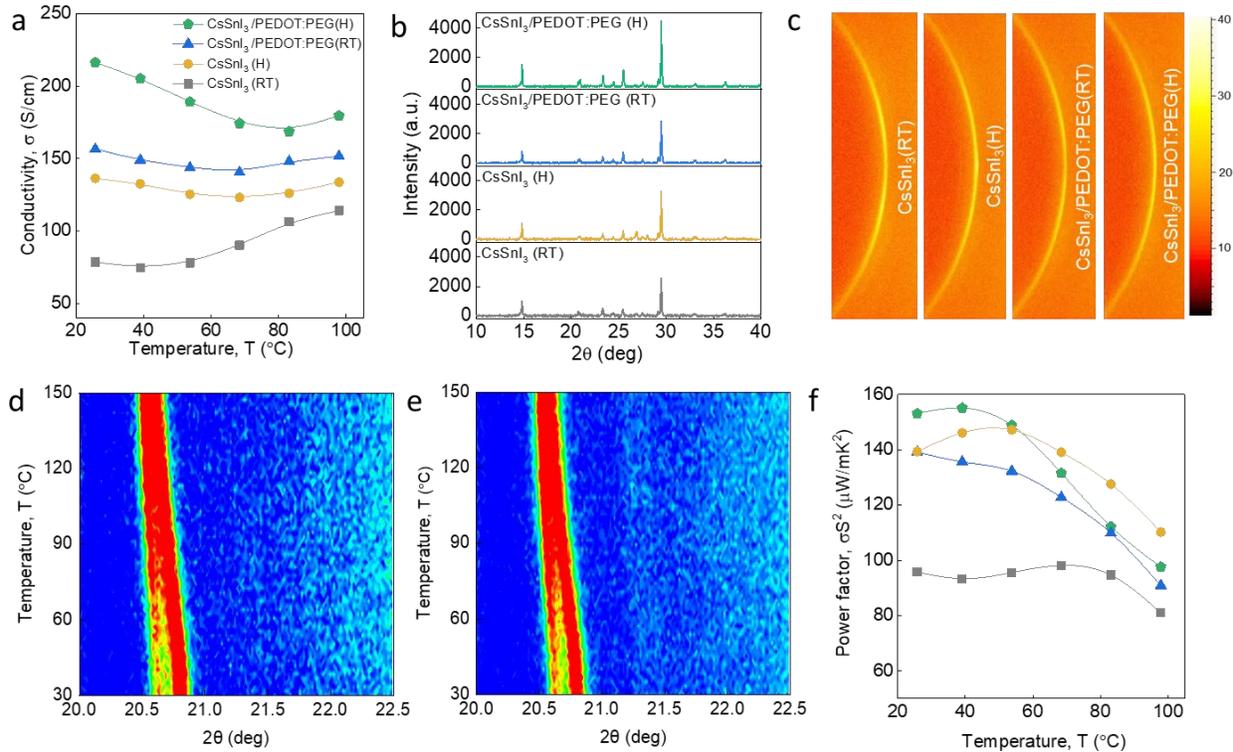

**Fig. 4 Process engineering and thermoelectric performance. a** Temperature-dependent electrical conductivity **b** XRD pattern **c** 2D XRD for RT and hot-cast CsSnI₃ and CsSnI₃/PEDOT:PEG films. In-situ temperature-dependent XRD showing the phase transition for **d** CsSnI₃(RT) and **e** CsSnI₃/PEDOT:PEG(H) films. **f** Thermoelectric power factor of RT and hot-cast CsSnI₃ and CsSnI₃/PEDOT:PEG films.

In summary, a series of organic modifiers were introduced into CsSnI₃ to manipulate its electrical properties. The incorporation of organic moieties impacted the electronic states of CsSnI₃, enabling high electrical conductivity along with tunability between semiconducting and metallic transport. This engineering of CsSnI₃ DOS with organic modifiers resulted in doping-like characteristics, leading to the simultaneous enhancement of carrier density and mobility. Furthermore, the p-and n-type thermometric performance demonstrated in this work are among the best for halide perovskites. The tunability offered by CsSnI₃ hybrids open up exciting possibilities for applications in other electronic devices, such as transistors and sensors.

**Table 1 Thermoelectric parameters of CsSnI₃(RT), CsSnI₃/PEDOT:PEG(H), and Cs₂SnI₆ films corresponding to the highest ZT.**

| Sample | Type | Temp. (°C) | $\sigma$ (S/cm) | S ($\mu$V/K) | K (W/mK) | PF ($\mu$W/mK$^2$) | ZT |
|---|---|---|---|---|---|---|---|
| CsSnI₃ | p | 68 | 90.1 | 104.3 | 0.28 | 98.1 | 0.12 |
| CsSnI₃/PEDOT:PEG | p | 53 | 189.1 | 88.7 | 0.23 | 148.8 | 0.21 |
| Cs₂SnI₆ | n | 97 | 1.41 | -295.6 | 0.12 | 12.3 | 0.038 |

**Methods**

**Materials**

SnI₂ (99.99%), CsI (99.999%), SnF₂ (99%), Oleylamine, PEDOT:PEG (0.7 wt% dispersion in Nitromethane), Polyaniline (emeraldine salt, short chain, grafted to lignin), Catechol, 2,5-Dihydroxy-1,4-benzoquinone DMF, DMSO, Chlorobenzene were purchased from Sigma-Aldrich and were used as received unless otherwise stated. MAI (Greatcell Solar), PbI₂ (99.9985% Alfa Aesar) P3HT (MW: 58-68K, Rieke metals), PEAI (>98%, TCI). PEDOT:PSS (AI 4083, Heraeus), PEG (Acros, MW: 300)

**Sample preparation:**

To prepare the CsSnI₃ precursor solution, CsI and SnI₂ with required molar ratios were dissolved in DMF and DMSO (a volume ratio of 8:2) with a total concentration of 1.3 M and stirred at room temperature. Pristine CsSnI₃ and CsSnI₃ with organic modifiers were prepared on glass by spin coating the precursor solution at 1000 rpm for 10 s and 4000 rpm for 30 s in an N₂ glove box. 100 µl Chlorobenzene was dripped on to the spinning substrate after the first 18 s of the spin-coating process. The spin-coated films were annealed at 50 °C for 20 min and then at 100 °C for 20 min. For all precursors, the volume ratio of DMF:DMSO was maintained as 8:2.

CsSnI₃/MAI: A solution of MAI of 4mg/ml in IPA was spin-coated on top of the already prepared CsSnI₃ films at 4000 rpm for 30 s. Then, films were annealed at 100 °C for 5 min.

CsSnI₃/Pb: 0.1 M PbI₂ was added to 1.2 M SnI₂ and 1.3 M CsI to prepare the perovskite solution.

CsSnI₃/OA: 0.2 wt.% of OA was added to the CsSnI₃ precursor solution.

$CsSnI_3/SnF_2$: 0.1 M $SnF_2$ was added to the $CsSnI_3$ precursor solution.

$CsSnI_3/OA/PEAI$: A PEAI solution of 4 mg/ml in IPA was spin-coated at 4000 rpm for 30 s on the surface of $CsSnI_3/OA$ films. Then, films were annealed at 100 °C for 5 min.

$CsSnI_3/P3HT$: 1 mg P3HT was dissolved in chlorobenzene and used as antisolvent.

$CsSnI_3/PD$, $CsSnI_3/PEG$, $CsSnI_3/CT$, $CsSnI_3/PANI$, $CsSnI_3/DB$: 1 mg/2 mg of PD, PEG, CT, PANI, DB were added to the $CsSnI_3$ precursor solution for $CsSnI_3/PD$, $CsSnI_3/PEG$, $CsSnI_3/CT$, $CsSnI_3/PANI$, $CsSnI_3/DB$ compositions, respectively. Phenazine-2,3-diol (PD) was synthesized according to literature procedures.[23]

$CsSnI_3/PEDOT:PSS$, $CsSnI_3/PEDOT:PEG$: PEDOT:PSS dispersion in water was drop casted on glass slide and heated to 100 °C until completely dry. Then it was scratched and dispersed in DMSO and ultra-sonicated for 15 minutes. Different volumes of PEDOT:PSS in DMSO was added to the $CsSnI_3$ precursor solution. First PEDOT:PEG was extracted from 10 ml of nitromethane by centrifugation and then washed and redispersed in 10 ml DMSO. Different volumes of PEDOT:PEG in DMSO were added to the $CsSnI_3$ precursor. For all $CsSnI_3$ precursor solutions, the DMF:DMSO volume ratio was maintained at 8:2 by adding required amount of pure DMSO.

$Cs_2SnI_6$: $Cs_2SnI_6$ films were obtained by leaving $CsSnI_3$ films in air for 24 h.

**Material characterization**

Scanning electron microscopy (SEM) images were obtained using FEI Nova Nano. X-ray diffraction patterns were obtained from the Bruker D8 ADVANCE diffractometer. UV-Vis absorption spectra were recorded using Perkin-Lambda spectrometer. NMR spectra were acquired using a Bruker 700 MHz AVANCE III NMR spectrometer equipped with Bruker CPTCI multinuclear CryoProbe (BrukerBioSpin, Rheinstetten, Germany).

**Thermoelectric measurements**

Electrical conductivity and Seebeck measurements were performed on Netzsch SBA 548 Nemesis thermoelectric set up under He environment. Fresh samples were measured by transferring from glove box to thermoelectric set up with less than 30 second air exposure. The instrument uncertainty for electrical conductivity and Seebeck measurements are ±5% and ±7%, respectively. In-plane thermal conductivity was measured under vacuum using the 3ω Völklein method with a Linseis Thin Film Analyzer. Samples were prepared onto pre-patterned test chips from Linseis. Charge carrier density was measured in vander pauw geometry on Lakeshore Hall system at RT under $N_2$ flow. Hole mobility was calculated by combining the electrical conductivity and charge carrier density using the expression $\mu = \sigma/ne$.[24]

## DFT methodology

The DFT calculations were performed using the FHI-aims all-electron code.[25-27] The default numerical settings, referred to as "intermediate" in FHI-aims were used. Local minimum-energy geometries of the Born-Oppenheimer surface were obtained with residual total energy gradients below $1 \times 10^{-2}$ eV $\text{Å}^{-1}$ for atomic positions by PBE-GGA functional[28] within the vdW correction following the Tkatchenko-Scheffler method[29] (PBE+TS). The band structure and density of states (DOS) are calculated following the PBE-GGA functional, including the spin-orbit coupling[30] with a k-point grid of $3 \times 3 \times 1$ to sample the Brillouin zone that corresponds to slabs shown in Figure 3.

## Abbreviations

0D: Zero-dimensional; 1D: One-dimensional; 2D: Two-dimensional; 3D: Three-dimensional; CT: Catechol; DB: 2,5-Dihydroxy-1,4-benzoquinone; DFT: Density-functional theory; DMF: Dimethylformamide; DMSO: Dimethyl sulfoxide; DOS: Density of states; LED: Light emitting diode; NMR: Nuclear magnetic resonance; PEDOT:PEG: Poly(3,4-ethylenedioxythiophene), bis-poly(ethyleneglycol), lauryl terminated; P3HT: Poly(3-hexylthiophene-2,5-diyl); PEDOT:PSS: Poly(3,4-ethylenedioxythiophene) polystyrene sulfonate; PEG: Polyethylene glycol; PD: Phenazine-2,3-diol; PANI: Polyaniline; RT: Room temperature; XRD: X-ray diffraction

## Acknowledgements

This publication is based upon work supported by the King Abdullah University of Science and Technology (KAUST) Office of Sponsored Research (OSR) under Award No. OSR-CRG2018-3737. Computations were performed on the Niagara supercomputer at the SciNet HPC Consortium. SciNet is funded by the Canada Foundation for Innovation; the Government of Ontario; Ontario Research Fund Research Excellence; and the University of Toronto.

## Author contributions

M.A.H and D.B designed the project. T.Z. performed the DFT calculations and wrote the analysis. M.A.H fabricated the samples, performed all thermoelectric measurements. L.H.H and R.T contributed to sample preparation. C.C synthesized the PD additive. B.D. performed 2D and temperature-dependent XRD measurements. A.H.E performed the NMR measurements. M.A.H wrote the first draft of the manuscript. E.H.S, F.P.G.A, D.B, and M.A.H reviewed and edited the manuscript.

**Competing interests**

The authors declare no competing interests.

**Data availability**

The main data supporting the findings of this study are available within the Article and its Supplementary Information.